\documentclass[prb,preprint,aps]{revtex4}
\usepackage{epsfig}
\newcommand{\beq}{\begin{eqnarray}} 
\newcommand{\eeq}{\end{eqnarray}} 
\newcommand{\bQ}{\bf Q}
\newcommand{\om}{\omega}
\newcommand{\be}{\begin{equation}}
\newcommand{\ee}{\end{equation}}

\newcommand{\bk}{{\bf k}}

\newcommand{\bq}{{\bf q}}
\newcommand{\s}{\sigma}

\newcommand{\lmb}{\lambda}
\newcommand{\pr}{^\prime}
\newcommand{\dpr}{^{\prime\prime}}

\newcommand{\ro}{\rho_0}
\newcommand{\roF}{\rho_0(\eps_F)}

\newcommand{\Gf}{Green function}
\newcommand{\Gfs}{Green functions}
\newcommand{\se}{self-energy}

\newcommand{\sgn}{\ {\rm sign}}
\newcommand{\nnn}{\nonumber\\}
\newcommand{\nn}{\nonumber}
\newcommand{\Jeff}{J_{\rm eff}}

\newcommand{\Lp}{L_{\phi}}
\newcommand{\hb}{\hbar}

\newcommand{\lp}{\left(}
\newcommand{\rp}{\right)}
\newcommand{\lb}{\left[}
\newcommand{\rb}{\right]}

\newcommand{\eps}{\varepsilon}
\renewcommand{\vec}[1]{{\mathbf{#1}}}

\begin{document} 
 
\title{The Kondo effect and weak localization}

\author{ Philip Phillips$^1$ and Ivar Martin$^2$}

%
\address{$^1$Loomis Laboratory of Physics\\
University of Illinois at Urbana-Champaign\\
1100 W.Green St., Urbana, IL, 61801-3080\\
$^2$Los Alamos National Laboratory, Los Alamos, NM. 87545}

%

\maketitle
\section{Introduction}

The word Kondo means battle in Swahili.  This coincidence is fortuitous because in the Kondo effect, a battle inevitably ensues anytime a magnetic impurity is
placed in a non-magnetic metal.  Below some energy scale,
the Kondo temperature ($T_k$) a lone magnetic impurity is robbed of its spin.  Above the Kondo temperature, rapid spin-flip scattering produces a temperature-dependent correction to the resistivity of the form, $B_k\ln T$. Until recently, both the Kondo resistivity and $T_k$ were thought to be determined solely
by the host metal and the magnetic impurity.  However, numerous presentations
in this volume attest, there is now overwhelming evidence that both are affected by the size of the sample\cite{s1,s2,s3,s4,s5} as well as
non-magnetic random scattering~\cite{d1,d2,d3,yanson}.   In this paper, I will focus on the theoretical work\cite{ivar}
we have performed on the experiments revealing that non-magnetic scattering
suppresses the Kondo resistivity in thin Kondo alloys. 

In Kondo alloys of the form Cu(Mn), Cu(Fe) and Au(Fe), Giordano and
 colleagues~\cite{d1,d2,d3} observed that introducing non-magnetic impurities suppressed the
coefficient of the Kondo logarithm.  The Kondo slope, $B_k$, is a monotonically
decreasing function as the mean-free path is decreased.  This result is surprising for two reasons. First, disorder gives rise to diffusive motion.  Hence,
relative to a clean sample, conduction electrons spend more time around
a given magnetic impurity in the presence of disorder.  Naively, this effect
would result in an enhancement of the 
Kondo resistivity.  Second, at the time of these experiments, the leading 
 theoretical view was that disorder eliminates the Kondo logarithm
and leads to a stronger algebraic divergence of the form $T^{d/2-2}$
in the resistivity. 
Everts and Keller\cite{everts} were the first to argue for the emergence
of a $1/\sqrt{T}$ in the Kondo
self-energy for a d=2 system in the presence of random non-magnetic scattering.
A few years lataer, Bohnen and Fisher\cite{bf} argued, however, that such a term would not survive
in the conductivity. More recently,
Ohkawa and Fukuyama\cite{ohk} and Vladar and Zimanyi\cite{vladar} have developed an extensive
diagrammatic scheme to re-investigate this problem and also concluded that
the algebraic singularity dominates the Kondo $\ln T$.  As a result,
these groups conclude that static disorder can mask the Kondo resistivity as
$T\rightarrow 0$.  The experiments show no singularity of this sort, however.
This complete lack of agreement between theory and experiment led us to re-evaluate the interplay between disorder and Kondo spin-flip scattering.   

As our work is based heavily on the previous diagrammatic expansion
of Fukuyama and colleagues~\cite{ohk}, it is first important to understand
how the algebaric divergence emerges from their analysis.  When non-magnetic impurities are present, the
diffusive propagator that describes the resultant motion
\beq
D(Q,\omega)\propto \frac {1}{(DQ^2-i\omega)}
\eeq
 has a diffusion pole.  Here,
$Q$ and $\omega$ are the net momentum and energy
transfer and $D=2\hbar\epsilon_F\tau/dm$. When such diffusive propagators are used to decorate the spin-flip vertices in the Kondo self-energy, the singular dependence found by Ohkawa and Fukuyama\cite{ohk} obtains as can be seen from the following
argument.   The most divergent contribution to the Kondo self-energy
arises from the two-diffuson decoration of the Kondo spin-flip vertices.
Diagrams of this form involve an integration over the internal momentum line:
\beq
\sum_{\bf Q} {\frac {1} {(D\bf Q^2 + |\omega|)^2}} \propto \int{\frac{Q^{d
      - 1}{dQ}}{(DQ^2 + |\omega|)^2}}\propto |\omega|^{d/2 -2}.
\eeq
The absolute value of the frequency appears here because we  work in the 
finite-temperature Matsubara formalism.  The Matsubara frequency
$\omega$ is proportional to temperature $T$.  Therefore, the temperature
dependence due to diffusons and Cooperons is indeed
 $T^{d/2 - 2}$, as can be also verified by a more careful
calculation~\cite{ohk}, and is a direct consequence of the diffusion
poles.  

The argument leading to the new algebraic dependence is certainly
clear.  However, it is well-known that spin-flip scattering can cut off the
 diffusion pole.  Should this occur then the algebraic dependence will only
be valid above a certain temperature, not as $T\rightarrow 0$.  Of course,
this requires that the feedback effect of spin-flip scattering on
localization physics be included.  It is this effect that has been
absent from all previous treatments of the disorder/Kondo problem.
Inclusion of the feedback effect of spin-flip scattering
on localization has been the primary focus of our work\cite{ivar}.
A key difference that the feedback effect introduces is a nontrivial
density dependence into the Kondo problem.  This difference
arises because diffusive propagators which include the spin-flip
scattering rate decorate the bare spin-flip vertices in the Kondo
self-energy.  The spin scattering
rate is proportional to the concentration
of magnetic impurities. Consequently, a non-zero spin-flip scattering rate arises only
if all the magnetic impurities are averaged over. Hence, the feedback effect
represents a departure from the single-impurity physics typically associated
with the Kondo problem.  That this state of affairs obtains naturally
when disorder is present can be seen from considering the standard weak-localization
correction\cite{alt}
\beq\label{weak}
\delta\sigma=-\frac{e^2}{2\pi^2\hbar}\ln\frac{\tau_\phi}{\tau_o}
\eeq
to the conductivity in a thin film, with $\tau_\phi$ the dephasing time
oand $\tau_o$ the elastic scattering time.  
Whenever localization physics is relevant, one has to decide
which is the dominant dephasing process.  Experiments show that the 
dephasing time is weakly dependent on temperature\cite{d1,d2,d3}.
This is consistent with a dephasing rate that is determined solely
by spin-flip scattering.  Hence,
$\hbar/\tau_s\propto n_s\hbar J^2>\hbar/\tau_T$,
where $\tau_T$ is the dephasing time due to all other processes in the 
system.  Consequently, if spin-flip scattering is the dominant
dephasing process, the number of impurities has a lower bound.  
Our treatment does not include impurity-impurity
effects, however.  What is crucial here is that the contribution from each
impurity must be averaged over to describe the dominant dephasing process.
Our central result that is used to compare with the experiments can be derived
simply from Eq.(\ref{weak}).  In the presence
of the Kondo logarithm, the spin-flip scattering rate is given by
\beq\label{kondo}
1/\tau_{s}=2/3\sqrt{3}\tau^o_s\left(1-4J_0N(0)\ln\frac{T_F}{T}+\cdots\right).
\eeq
Substitution of this result into Eq.~(\ref{weak}) and expansion
of the logarithm for $T>T_k$ yields
 the contribution of spin-flip scattering to the conductivity
\beq\label{wl}
\delta\sigma\approx \sigma_0\left(\ln\frac{\tau_s^0}{\tau_o}
-N(0)J_0\ln\frac{T_F}{T}\right)
\eeq
for a $d=2$ sample, with $\sigma_0$ the Drude conductivity.
Because $J_0<0$, the Kondo logarithmic term enhances the
spin-scattering time and in turn reduces the magnitude of the 
weak-localization correction.  That is, spin-flip scattering produces an `antiloclization' effect.  Further, this correction is opposite in sign to the zeroth-order
Kondo logarithm.  Consequently, disorder leads to a suppression of the Kondo
resistivity.  The suppression of the Kondo resistivity follows immediately from three principles: 1) spin-flip scattering feeds back into the Kondo self-energy
to regularize the algebraic divergence, 2) weak localization appears as a negative correction to the conductivity and 3) spin-flip scattering weakens the weak-localization effect. Hence, the net effect is a positive correction logarithmic correction to the conductivity which when added to the negative bare Kondo logarithm leads to a diminished logarithmic conductivity.

\section{Formulation of Problem}
 
The starting point for our analysis is a model Hamiltonian $H=H_o+H_{sd}$ 
that contains both normal impurities
\begin{equation}
H_o=\sum_{k\sigma}\left(\varepsilon_k-\varepsilon_F\right)a_{k\sigma}^{\dagger}a_{k\sigma}
+{\frac v\Omega}\sum_{k,k',i}e^{\bf i(k-k^{\prime})\cdot R_i}
a^{\dagger}_{k\sigma} a_{k'\sigma}
\end{equation}
as well as magnetic scatterers
\beq
H_{sd}=-{\frac J \Omega}\sum_{R_n,k,k',\sigma,\sigma'}
e^{\bf i(k-k^{\prime})\cdot R_n}{\bf \sigma}_{\sigma,\sigma'}\cdot {\bf S_n}
a_{k\sigma}^{\dagger}a_{k'\sigma'}.
\eeq
where $v$ measures 
the strength of the 
scattering with the non-magnetic disorder, $R_n$ 
denotes the position of the impurities, magnetic or otherwise, ${\bf S}_n$ is the spin
operator for the magnetic impurity at site $n$, and $\Omega$ is the volume. 
 The two natural timescales in this problem
are, $\tau_s^o$ and $\tau_o$, the bare magnetic and non-magnetic scattering times.
In terms of the density of states of the host metal, $\rho_o$ and the 
concentrations of magnetic and non-magnetic scatterers, $n_s$ and $n_o$, respectively,
we have that $\hbar/2\tau^o_s=3\pi n_s\rho_o |J|^2/4$ and 
$\hbar/2\tau_o=\pi n_o\rho_o |v|^2$. The total scattering rate
is $1/\tau=1/\tau^o_s+1/\tau_o$.   To measure the strength of the non-magnetic
disorder, we define $\lambda=\hbar/(2\pi\varepsilon_F\tau_o)$.  We assume that the concentration
of localized spins is dilute so that long-range spin glass effects are irrelevant.
Also, we work in the regime in which normal impurity scattering dominates, $1/\tau_o\gg 1/\tau_s^o$.  

Describing  scattering in the presence of a weakly disordered
potential requires Cooperon and diffuson propagators.  The traditional
form of such propagators, $C(Q,\omega)=D(Q,\omega)\propto (DQ^2-i\omega)^{-1}$,
was used extensively in the early treatments\cite{ohk} of the disordered Kondo problem.
However, as remarked in the introduction, such a procedure assumes that 
diffusive motion with a diffusion pole remains intact even in the presence of oscillating fields created by spin-flip scattering. It is this assumption that leads to the divergence found earlier by Fukuyama and co-workers\cite{ohk}.  To
alleviate this problem, we include the all-important feedback effect spin-flip scattering
has on such diffusive processes.  If all scattering processes are treated
in the first Born approximation, the Dyson-like integral equation,
\beq\label{diffu}
D_{\alpha \beta \delta \gamma}=\delta_{\alpha \beta} \delta_{\delta
\gamma}+ \overline {U_{\alpha \mu} U_{\nu \gamma}} \sum G_\mu^R
G_\nu^A D_{\mu \beta \delta \nu}
\eeq
describes all ladder diagrams with the spin-dependent potential
\beq
{\overline {U_{\alpha \mu} U_{\nu \gamma}}= \frac{1}{\tau_o}
\delta_{\alpha \mu} \delta_{\nu \gamma}+ \frac{1}{3 \tau_s}
{\vec{\sigma}}_{\alpha \mu} \cdot {\vec{\sigma}}_{\nu \gamma}}.
\eeq
The Greek letters denote the spin indices on the upper and lower electron lines
in the diffuson ladder and repeated indices are summed over.
The advanced and retarded Green functions are given by
$\left(G_\mu^A\right)^{-1}=\epsilon_F - p_-^2/2m - i/2\tau +
\nu h$
and $\left(G_\mu^R\right)^{-1}=\epsilon_F + \omega - p_+^2/2m + i/2\tau + 
\mu h$ respectively.  Noting that 
\beq
\sum G_\mu^R G_\nu^A = - \sum
\frac{G_\mu^R-G_\nu^A}{(G_\mu^R)^{-1}-(G_\mu^A)^{-1}}
\eeq
we arrive at the solution for the diffuson,
\beq\label{eq:dif}
D_{\alpha\beta\gamma\delta}&=&\frac {\hbar}{4\tau (DQ^2-i\omega)}
\left(\delta_{\alpha\beta}\delta_{\gamma\delta}+\sigma_{\alpha\beta}\cdot\sigma_{\gamma\delta}\right)
\nonumber\\
&+&\frac{\hbar}{4\tau (DQ^2-i\omega+4/3\tau_s^0)}\left(3\delta_{\alpha\beta}
\delta_{\gamma\delta}-\sigma_{\alpha\beta}\cdot\sigma_{\gamma\delta}\right).
\eeq
The analogous integral equation for the Cooperon
\beq
C_{\alpha \beta \gamma \delta}=\delta_{\alpha \beta} \delta_{\delta
\gamma}+ \overline {U_{\alpha \mu} U_{\gamma \nu}} \sum G_\mu^R
G_\nu^A C_{\mu \beta \nu \delta}
\eeq
can be solved analogously to yield,
\beq
C_{\alpha\beta\gamma\delta}&=&\frac {\hbar}{4\tau (DQ^2-i\omega+2/\tau_s^0)}
\left(\delta_{\alpha\beta}\delta_{\gamma\delta}-\sigma_{\alpha\beta}\cdot\sigma_{\gamma\delta}\right)
\nonumber\\
&+&\frac {\hbar}{4\tau(DQ^2-i\omega+2/3\tau_s^0)}
\left(3\delta_{\alpha\beta}\delta_{\gamma\delta}+\sigma_{\alpha\beta}\cdot\sigma_{\gamma\delta}\right).
\eeq
where $\alpha\beta$ and $\gamma\delta$ are spin indices.  The
dot-product $\sigma_{\alpha\beta}\cdot\sigma_{\gamma\delta} =
\sigma^x_{\alpha\beta} \sigma^x_{\gamma\delta} + \sigma^y_{\alpha\beta}
\sigma^y_{\gamma\delta} + \sigma^z_{\alpha\beta}
\sigma^z_{\gamma\delta}$. 

\section{Self Energy}

\begin{figure}[htbp]
  \begin{center}
    \includegraphics[width = 5 in]{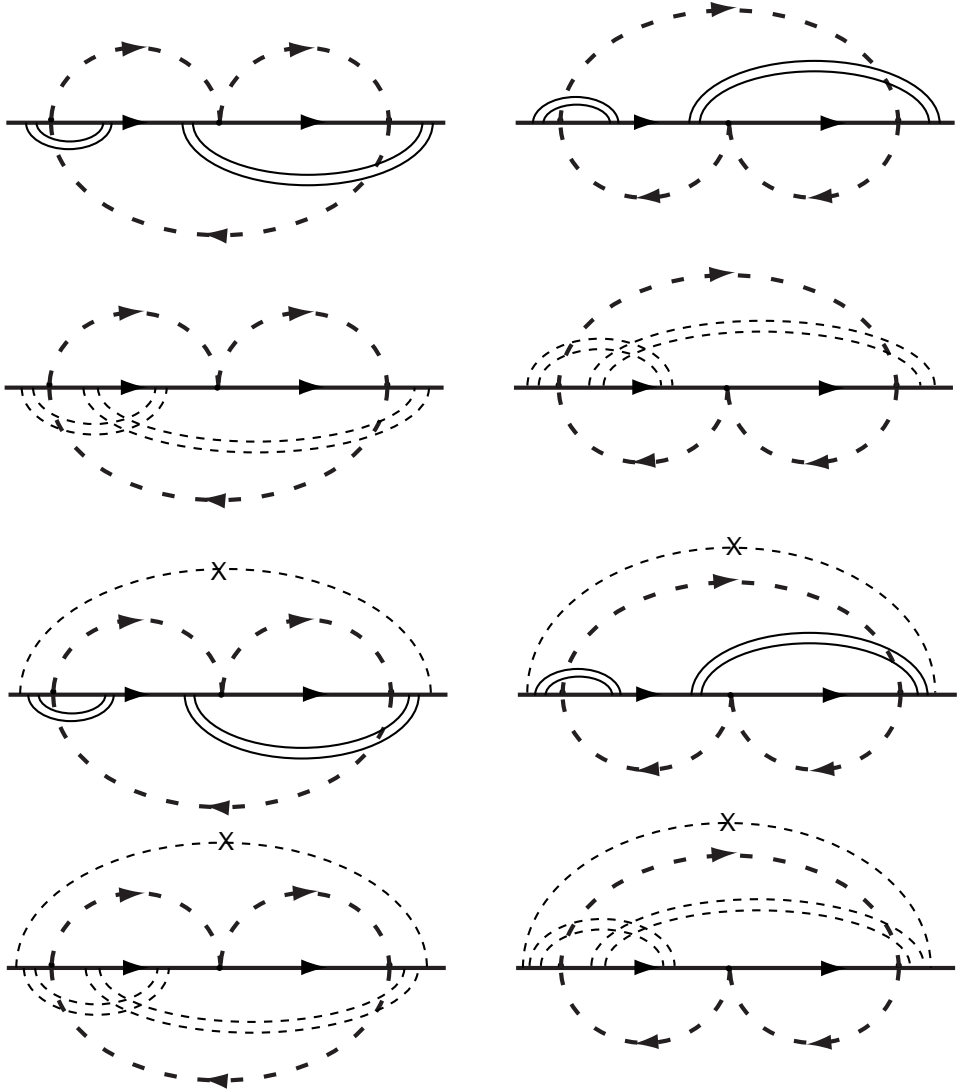}
    \caption{ Feynman diagrams contributing to the Kondo self-energy.
      The dashed lines correspond to Abrikosov
      pseudofermions and the  double solid 
      lines to diffusons and double dashed lines to the Cooperons.  The Greek
      letters indicate the spin. The $X$ indicates a single
      non-magnetic impurity scattering event.  Such diagrams are known
as the rainbow diagrams.} 
    \label{fig:16}
  \end{center}
\end{figure}

As is evident, even in the presence of spin-flip scattering, the diffuson
still retains its diffusion pole in the $S=0$ channel.  Hence, we will be back
to where we started if the $S=0$ diffuson still contributes to the 
self-energy.  We now show that this contribution vanishes identically to 
all orders of perturbation theory.  Consider the self-energy diagrams 
shown in Fig.(\ref{fig:16}).

To illustrate how the self-energy diagrams in Fig.~\ref{fig:16} are evaluated
let's focus on the first two  diagrams with the diffuson vertex decorations.
The sum of the two self-energy diagrams
is  
\beq
\label{eq:selfD}
\Sigma_{3q}^D(\bk,i\epsilon_n)&=&   n_s J^3 T\sum_{\omega_\ell,\omega_m,\bQ,\bq}
\Theta(-  \epsilon_n(\epsilon_n+\omega_\ell))
V_{\alpha\beta\nu\eta}(i\omega_\ell,i\omega_m) \nnn
&&\times G(i\epsilon_n+i\omega_m,\bq) G(i\epsilon_n+i\omega_\ell,\bk+\bQ)\nnn
&&\times D_{\sigma\alpha\beta\gamma}(i\omega_\ell,\bQ) 
D_{\gamma\nu\eta\sigma}(i\omega_\ell,\bQ)
\eeq
where $G(i\epsilon,q)$ is the electron Green function
\beq
G(i\epsilon,q)=\frac{1}{i\epsilon+\epsilon_F-\hbar^2q^2/2m+i(\hbar/2\tau)
{\rm sign}(\epsilon)},   
\eeq
and the electron energies are given by
the Matsubara frequencies, \mbox{$\epsilon_n=(2n+1)\pi T$}.   The pseudofermion
energies are defined in terms of $z_k=(2k+1)\pi T$ and
$\omega_\ell=2l\pi T$.
The range of summation over the momentum $\bQ$ and energy $\om_\ell$
transfers is limited by the range of validity of the diffusion
approximation, $DQ^2<\hbar/\tau$ and $\om_\ell < \hbar/\tau$.  The
step function $\Theta(x)$ appears in the expression because the
diffusion propagators are only non-zero if the impurity ladders connect
electrons on different sides of the Fermi surface.  
The summation over momenta $\bk\pr$ and $\bk\dpr$ in the \Gfs\
adjacent to the spin vertices is already included in the definition of
the diffuson.
The \Gfs\ in Eq.~(\ref{eq:selfD}) can be simplified using
\beq
\label{eq:simp}
\sum_{\bq}G(i\epsilon_n+i\omega_m,\bq) &\approx& - i\pi\roF \sgn(\epsilon_n
+ \omega_m), \nnn
G(i\epsilon_n+i\omega_\ell,\bk+\bQ) &\approx& -i
\frac{2\tau}{\hb}\sgn(\epsilon_n+\omega_\ell) =
i\frac{2\tau}{\hb}\sgn(\epsilon_n). 
\eeq
The first approximation can be obtained by integration around the
Fermi surface. The second approximation makes use of the fact that
the momentum $\bk$ and energy $\epsilon_n$ are close 
to the Fermi surface (within the energy shell of width $T$), and the
momentum and energy transfers allowed by the diffusion propagator are
less than $\hb/\tau$.

The pseudofermion part, 
\beq
\label{eq:VpfS}
V_{\alpha\beta\nu\eta}(\om_\ell, \om_m) = -\frac{1}{16}
\left[\frac{\delta_{m 0}}{i\om_\ell}(1-\delta_{\ell 0})
  +\frac{\delta_{\ell 0}}{i\omega_m}(1-\delta_{m 0})
  -\frac{\delta_{\ell m}}{i\om_\ell}(1-\delta_{\ell
    0})\right](\sigma_{\alpha\beta}\cdot \sigma_{\nu\eta}). 
\eeq
involves a trace over the
impurity spin states. The internal spin indices are not summed over
because the electron  spin can be flipped 
by the diffusons. 
After substituting all the ingredients into Eq.~(\ref{eq:selfD}) and performing
the summation over the spin indices, the self-energy becomes
\beq
\Sigma_{3q}^D(\bk,i\epsilon_n) = - 6 n_s \pi \ro J^3 \tau T \sum_{\bQ,
  \om_\ell} {\frac {\Theta(-
    \epsilon_n(\epsilon_n+\omega_\ell))}{i\om_\ell} \lb
  \frac{\hb/\tau}{D\bQ^2 + |\om_\ell| + 4\hb/3\tau_s^0}\rb^2}
\sgn(\epsilon_n).\nn 
\eeq
Remarkably, the divergent $S = 0$ part of the diffuson drops out, and
as a result the singular temperature dependence in the resistivity
disappears. 

But what is the source of this cancellation and is it exact?  By careful examination of the pseudofermion contribution
Eq.~(\ref{eq:VpfS}), we see that the sum over 
the spin indices in the self-energy [Eq.~(\ref{eq:selfD})] separates
into two identical sums of the form, 
$\sum_{\alpha\beta}D_{\sigma\alpha\beta\gamma}
\sigma_{\alpha\beta}^a$.
If we use the identity
$\sum_{\alpha\beta}(\sigma_{\nu\alpha}\cdot\sigma_{\beta\gamma})\cdot
\sigma_{\alpha\beta}^a = -\sigma_{\nu\gamma}^a$,
we find immediately that the cancellation of the $S=0$ diffuson
\beq\label{cancel}  
\sum_{\alpha\beta}D^{S=0}_{\nu\alpha\beta\gamma}\sigma_{\alpha\beta}^a
&\propto& \sum_{\alpha\beta}(\delta_{\nu\alpha}
\delta_{\beta\gamma}+\sigma_{\nu\alpha}\cdot\sigma_{\beta\gamma})
\sigma_{\alpha\beta}^a=0
\eeq
from the $3^{rd}$ order Kondo self-energy is exact. To any order in $J$
in the most divergent approximation, the cancellation
of the $S=0$ diffuson can be seen as follows.  Within
this scheme, each diffuson encircles a vertex that is exactly
equal to the Abrikosov~\cite{abr1965} vertex function 
$\Gamma = \Jeff\sigma\cdot {\bf S}$, with $\Jeff$ defined
within the parquet summation.  When this function 
is now multiplied by $D^{S=0}$ and summed over the spin indices,
the cancellation to all orders follows immediately from Eq. (\ref{cancel}).
Note that this cancellation relies on the spin algebra and hence is
not tied to 
the approximations used to obtain $D^{S=0}$. 
Consequently, the cancellation of the $S=0$ component of the diffuson 
is fundamentally tied to the fact that the Kondo
interaction does not conserve the electron's spin. This cancellation
theorem which signifies that the Kondo logarithm remains in tact
 is in the spirit of Anderson's theorem that non-magnetic impurities
do not affect $T_c$ for s-wave superconductors.  Recently,
Chakravarty and Nayak\cite{chak} have shown that in the very weak-disorder
limit, a true Anderson theorem exists in which disorder does scale out of
the Kondo problem.

As advertised, the cancellation of the diffusion pole suppresses the 
algebraic divergence of the self-energy.  To see how this emerges, we continue with our analysis of the first two diagrams in Fig.(\ref{fig:16}).
We limit ourselves to 2D case, and hence
\beq
\Sigma_{3q}^D(\bk,i\epsilon_n) &=&  6 i n_s \pi^2 \ro^2 J^3 \lmb
\frac{\hb}{\tau} T \sum_{ 
  \om_\ell} {\frac {\Theta(-\epsilon_n(\epsilon_n+\omega_\ell))}
  {\om_\ell (|\om_\ell| +  4\hb/3\tau_s^0)}} \sgn(\epsilon_n)\nnn
  &=& i \frac 32  n_s  \ro^2 J^3 \lmb \frac{\hb}{\tau T}  \sum_{
  m = 0 } ^\infty {\frac {1}{\lp m + \frac{\eps_n}{2\pi T} + \frac
    12\rp 
    \lp m + \frac {\eps_n}{2\pi T} + \frac 1 2 +
    \frac {2\hb}{3\pi T \tau_s^0} \rp}} \sgn(\epsilon_n).\nn
\eeq
where $\lmb$ is the dimensionless disorder defined earlier.
From this self energy, we define the scattering rate as follows:
\beq
\frac \hb {2 \tau^{q}_D} &=& \int{\lp - \frac{\partial f}{\partial \epsilon} \rp
  (-{\rm Im}\ \Sigma_{3q}^D(\bk,\epsilon + i 0)) d \epsilon}\nnn 
&=& - \int {f(\epsilon) \frac{\partial {\rm Im}\ \Sigma_{3qD}(\bk,\epsilon +
  i 0)}{\partial \epsilon} d \epsilon}.\nn
\eeq
Clearly, one cannot evaluate this expression just by setting $\epsilon
 = 0$ because of the singular temperature dependence in the \se.  We will 
compute
 this expression  by contour integration in the complex
 $\epsilon$-plane.  The self-energy has two poles in the upper
 half-plane.  Hence if we close the contour in the lower half-plane,
 then the integral will be equal to the sum of the residues in the
 points where the Fermi function $f(\epsilon)$ has poles, 
 $\epsilon_k = - i (2k +1) \pi T$:
\beq
\frac \hb {2 \tau^{q}_D} 
&=& 2\pi i T  \sum_k {\frac{\partial {\rm Im}\ \Sigma_{3qD}(\bk,\epsilon +
  i 0)}{\partial \epsilon}\Big|_{\epsilon = \epsilon_k}}\nnn
&=& \sum_{k,m = 0}^\infty{\frac{1}{(m + k + 1)(m + k + 1 + \eta)}
  \lb \frac{1}{(m + k + 1)} + \frac{1}{(m + k + 1 + \eta)} \rb}\nn
\eeq
where $A$ is the coefficient of the self-energy and $\eta = 2\hb/3\pi
T \tau_s^0$.  In the double sum over $m$ and $k$ there are $(m + k +
1) \equiv N$ identical elements.  Therefore, the sum can be
transformed to a sum over $N$ times $N$.  Now if we use the series
expansion for the digamma, $\psi(x) = d\ln \Gamma(x)/dx$, and
trigamma, $\psi\pr(x) = d^2\ln \Gamma(x)/dx^2$ functions,
\beq
\psi(1+x) &=& -\gamma + z \sum_{N = 1}^{\infty} \frac{1}{N(N + x)}\nnn
\psi\pr(1+x) &=& \sum_{N = 1}^{\infty} \frac{1}{(N + x)^2},\nn
\eeq
and define a new function 
$$ F(x) \equiv \frac{\psi(1 + x) + \gamma}{x} + \psi\pr(1 + x)$$
then the result for the contribution to the scattering rate becomes
\beq
\label{eq:tqD}
\frac \hb {2 \tau^{q}_D} = - \frac 32  n_s  \ro^2 J^3 \lmb \frac{\hb}{\tau
  T} F\lp \frac {2\hb}{3\pi T \tau_s^0}\rp.
\eeq
For small argument, $x\ll 1$, $F(x) \approx \zeta(2) = \pi^2/6$; for
large argument $x \gg 1$, $F(x) \approx \ln (x)/x$.  Hence we
conclude that there are two regimes, corresponding to ``high''
($\hb/\tau_s^0 > T$) and ``low'' ($\hb/\tau_s^0 < T$) impurity
concentrations, in which the diffuson corrections behave
logarithmically in temperature and as $1/T$, respectively.
In other words, the $1/T$ behavior is cut-off at the temperatures below
$\hb/\tau_s^0$.
Now let us consider the rest of the diagrams, namely the diagrams that
involve Cooperon propagators, and the diagrams with the external
single impurity line.
\beq\label{eq:selfC}
\Sigma_{3q}^D(\bk,i\epsilon_n)&=&  n_s J^3 T\sum_{\omega_\ell,\omega_m,\bQ,\bq}
\Theta(-  \epsilon_n(\epsilon_n+\omega_\ell))
V_{\alpha\beta\nu\eta}(i\omega_\ell,i\omega_m) \nnn
&&\times G(i\epsilon_n+i\omega_m,\bq) G(i\epsilon_n+i\omega_\ell,\bk+\bQ)\nnn
&&\times C_{\sigma\alpha\gamma\nu}(i\omega_\ell,\bQ) 
C_{\beta\gamma\eta\sigma}(i\omega_\ell,\bQ).
\eeq
The only difference compared to  the  diffuson self-energy 
contribution is the different spin indexing of the Cooperons compared
to the diffusons.  This is because the Cooperon propagators need to be 
``crossed'' in order to have the same momentum transfers without
phase space restrictions.
Summing over the spin indices reduces the problem to one in which the
product of the  Cooperons is spin independent and equal to:
\beq
\tilde{C}^2&=&\frac{\hbar^2}{2\tau^2}
\left[\frac{1}{(DQ^2-i\omega+2/\tau_s^0)^2}+   
\frac{1}{(DQ^2-i\omega+2/3\tau_s^0)^2}\right].   
\eeq
Now we can continue in exactly the same way as in  the diffuson case
to obtain for the scattering rate correction
\beq
\label{eq:tqC}
\frac \hb {2 \tau^{q}_C} = - \frac 32  n_s  \ro^2 J^3 \lmb \frac{\hb}{\tau T} 
\lb
   \frac 1 2 F\lp \frac {\hb}{\pi T \tau_s^0} \rp
+  \frac 1 2 F\lp \frac {\hb}{3\pi T \tau_s^0} \rp 
\rb.
\eeq
At high temperatures, $T > \hb/\tau_s^0$ we again recover the $1/T$
behavior, and for low temperatures, $T < \hb/\tau_s^0$, the logarithmic
behavior obtains. 

Finally, we need to consider the set of
diagrams that contain one external impurity line (the rainbow diagrams). 
We will prove now
that such diagrams are equal to the corresponding diagrams without the
impurity line times the factor of (-1/2).  Unlike the diagrams that we
considered before, the internal \Gf\
$G(i\epsilon_n+i\omega_\ell,\bk+\bQ)$ can no longer be replaced by its
value at the Fermi surface, $2 i \tau/\hb$.  Instead, a sum over the
intermediate momentum of a product of three \Gfs\ needs to be computed:
\beq
|v|^2 \sum_{\bk\pr}{\frac{1}{(i\epsilon_n + \epsilon_F - \epsilon_{\bk\pr} +
    i\hb/2\tau)^2(i\epsilon_n +i\om_\ell + \epsilon_F -
    \epsilon_{\bk\pr + \bQ} - i\hb/2\tau)}}\nnn
\approx |v|^2 \int_{\infty}^{-\infty}{\frac{\ro d(-x)}{(x + i \hb/2\tau)^2(x -
    i\hb/2\tau)}} = 2 \pi i \ro |v|^2 \frac{1}{(2 i \hb /2\tau)^2} =
-i \tau/\hb. 
\eeq
In deriving this relation, we set $\bQ$ and $\om_\ell$ to zero, since
they are small.  Therefore the  sum of the diagrams with and without the
external impurity lines is two times smaller that sum of the diffuson
and Cooperon diagrams that we derived before.  However, there is a
factor of 2 that comes from two possible internal electron lines to
which the diffusion propagators can be attached.  As a result, the
sum of {\em all} diagrams happens to be exactly equal to the sum of
two contributions that we already computed.  Hence the total quantum
correction to 
the scattering time due to the Kondo diagrams in Fig.~\ref{fig:16} is 
\beq
\frac{1}{\tau^q} = \frac{1}{\tau^{q}_D} + \frac{1}{\tau^{q}_C},\nn
\eeq
with $1/\tau^{q}_D$ and $1/\tau^{q}_C$ given by Eq.~(\ref{eq:tqD}) and
Eq.~(\ref{eq:tqC}), respectively.

\section{Conductivity}

The total conductivity is a sum
of the Drude contribution with the transport scattering time, and
the weak localization  correction.  The transport scattering rate is
composed of elastic scattering, the second order in $J$ spin
scattering, the third order in $J$ (Kondo) scattering and the quantum
corrections computed above. 
\beq
\frac{1}{\tau_{tr}} = \frac{1}{\tau_{0}} + \frac{1}{\tau_s^{0}} +
\frac{1}{\tau_s^{K}} + \frac{1}{\tau_{D}^q}  + \frac{1}{\tau_{C}^q}.
\eeq
Of these contributions, only the last three have non-trivial
temperature dependence.  The total conductivity is
\beq
\sigma(T) = \frac{e^2 n \tau_{tr}} {m} + \delta\s_{WL} = \s_0\lp 1 -
\frac{\tau}{\tau_s^K} - \frac{\tau}{\tau_D^q}- \frac{\tau}{\tau_C^q} +
\frac{\delta\s_{WL}}{\s_0}\rp, 
\eeq
where $\s_0$ is the temperature-independent part of the conductivity.
There are also more complex conductivity diagrams that involve both the
spin-dependent pseudofermion part and the diffusion propagators, but
they can be shown to cancel out~\cite{ohk}. 

The derived expressions for $\tau_{D}^q$ and $\tau_{C}^q$ have simple
asymptotic behavior.  
For $d=2$ in the limit $T\gg \hbar/\tau^0_S$, we recover the inverse
temperature dependence 
\beq\label{hight}
\frac{\hbar}{2\tau^{C}}=\frac{\hbar}{2\tau^{D}}\approx
\frac{-\pi\hbar\rho_0\lambda J}{3\tau}\frac{\hbar}{\tau_s^0
  T}\ll-\rho_0\lambda J\frac{\hbar}{\tau} 
\eeq
of Refs. (\cite{vladar,ohk}).  Without the diffusion pole
cancellation, the lower bound in temperature for the $1/T$ behavior
would be set by ${\rm max}[{\hbar/\tau_{\phi},T_K}]$,
where $\tau_{\phi}$ is the inelastic scattering time. We find here that by 
explicitly including spin-scattering in the diffusion propagators, the
algebraic behavior occurs when $\hbar/(\tau_s^0 T)\ll 1$.  We will
see later that as a result of this  restriction, 
the contribution of the $1/T$ term to the conductivity is
negligible. In the opposite regime, $T\ll\hbar/\tau^0_S$, the
scattering rates 
\beq\label{lowt1}
\frac{\hbar}{2\tau^{D}} &=& -\frac 3 2\rho_0\lambda J 
\frac{\hbar}{\tau}\ln\frac{\hbar}{T\tau_s^0}\nnn
\frac{\hbar}{2\tau^{C}}&=&- 2\rho_0\lambda J 
\frac{\hbar}{\tau}\ln\frac{\hbar}{T\tau_s^0}
\eeq
are both logarithmic functions of temperature.   

The weak-localization contribution is given in Eq.(\ref{wl}. We collect all the contributions discussed above to determine the
temperature-dependent conductivity.
In the temperature range $T_K< T<\hbar/\tau_s^0$,  Cooperon,
diffuson, and weak-localization 
corrections are all logarithmic in temperature. Combining the results
from Eq.~(\ref{lowt1}) with the weak-localization correction, we find  
that the magnitude of the logarithmic part of the conductivity 
\beq
\sigma^T=\sigma_0\frac{4\tau \rho_0 J}{\tau_s^0}\left(1 +
  0.75\lambda\frac{\tau_s^0} 
{\tau}\right)\ln\frac{\epsilon_F}{T}.
\eeq
The first term in this expression arises from the unperturbed
Kondo effect and the latter from the interplay with disorder.  
Inclusion of disorder in the self-energy, even after inclusion of the
negative WL correction, enhances the Kondo
resistivity relative to a clean system result.

For temperatures $T\gg \hbar/\tau_s^0$, the self-energy
contribution to the relaxation time scales as $1/T$, whereas the 
weak-localization
correction is proportional to $\ln T$. However, comparison of the magnitude
of these corrections
reveals that the weak-localization term dominates, and the magnitude of
the resultant temperature-dependent conductivity
\beq\label{supps}
\sigma^T=\sigma_0\frac{4\tau\rho_0 J}{\tau_s^0}\left(1-
\frac{\lambda\tau_s^0}
{\tau}\right)\ln \frac{\eps_F}{T}
\eeq
is suppressed by the disorder.
Let us now apply our results to thin films with a thickness, $L$. 
We are interested in thin films, such that $\ell<L\ll L_\phi$.
Because
$\ell<L$, the electron gas is characterized by a 3-dimensional 
density of states
$\rho_0=1/(2\pi)^2(2m/\hbar^2)^{3/2}\epsilon_F^{1/2}$ and diffusion 
constant given by $D=2\hbar\epsilon_F\tau/3m$. 
Since the dephasing length $\Lp$ exceeds the film thickness, such a film
should be treated as quasi-2D with respect to weak localization.  That
means that the momentum-transfer summation in the diffusion
propagators must be restricted to the plane, or $\sum_{\bQ}
\rightarrow (1/L)\sum_{\bQ (2D)}$. The density of states
that arises from converting this sum into an integral is the
two-dimensional 
density of states,  $\rho_0^{2D}=\pi\rho_0/k_F$.  Hence, the self-energy
diagrams with the diffusion propagators will generate a
size-dependence to the conductivity of the form $1/(k_F L)$. 
The explicit finite-size weak-localization correction is~\cite{volkov}
$$\delta\sigma_{WL}=-\frac{e^2}{2\pi^2\hbar L}\ln  
\left(\frac{3 \sqrt 3\tau_s}{2\tau} \sinh\lp\frac L \ell\rp\frac  
    \ell L\right)$$. 
The size-dependence under the logarithm yields an effective size
dependence in 
the spin-relaxation time.  This size dependence should be observable 
in the standard WL  magnetoresistance measurements in the weak
magnetic fields.   However, it will not affect the temperature dependence 
of the conductivity.  The only size dependence that is coupled to the
temperature is the $1/L$ prefactor of the weak-localization correction.

We now combine these results in the low and high-temperature limits discussed
earlier.  In the two limits, we obtain conductivities
\begin{equation}\label{final}
\sigma^T=\left\{\begin{array}{ll}
\sigma_0\frac{4\tau\rho_0 J}{\tau_s^0}\left(1+
\frac{0.25\hbar\tau_s^0}
{ m k_FL\ell^2}\right)\ln \frac{\epsilon_F}{T}&{\rm if}\quad
T_K\ll T<\hbar/\tau_s^0\\ 
\sigma_0\frac{4\tau\rho J}{\tau_s^0}\left(1- 
\frac{1.5\hbar\tau_s^0}
{mk_FL\ell^2}\right)\ln \frac{\epsilon_F}{T}&{\rm if}\quad
T_K,\hbar/\tau_s^0\ll T \end{array}\right. 
\end{equation}
that have an explicit size and disorder correction that scales as
$1/(\ell^2L)$. The 
fact that only the coefficient of $\ln T$, but not the form of the
temperature dependence,  is modified is a direct consequence
of the diffusion pole cancellation theorem.  When magnetic
impurity density is high, we find and enhancement of the Kondo logarithm.
This is an intuitive result since, qualitatively, diffusive motion of
electrons is 
expected to enhance the probability of repeated scattering that
generates the Kondo effect.  The surprising finding is that in the
other regime, an overall suppression of the logarithmic correction in
the conductivity is obtained.  
While the self-energy enhancement is always present,
as it can be seen from the positive self-energy corrections to the
transport scattering rate [Eqs.~(\ref{eq:tqD}) and (\ref{eq:tqC})], this
effect is completely overwhelmed in the conductivity by the
WL correction 
which also acquires $\ln T$ dependence due to the Kondo contribution
to the dephasing rate.

\section{Experimental Applications}

\begin{figure}[htbp]
  \begin{center}
    \includegraphics[ height= 5 in, width = 3 in, angle = -90]
    {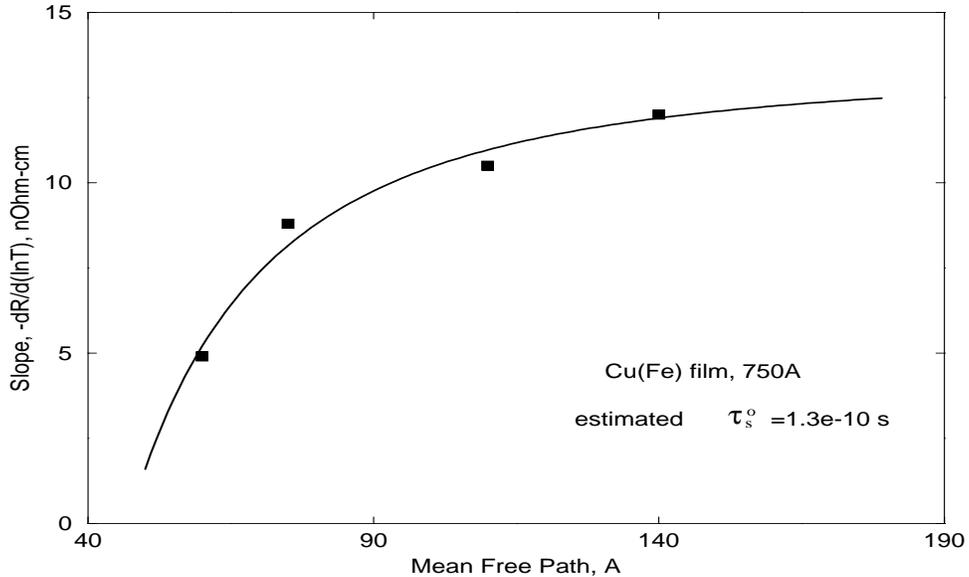} 
    \caption{Comparison of  the theoretical
      Kondo resistivity predicted from the second 
      of Eqs. (\protect\ref{final}) with the experimental data of 
      Blachly and Giordano.  
      The horizontal axis measures the strength of
      the static disorder through the mean-free path.}
    \label{fig:comp}
  \end{center}
\end{figure}

In the experiments of Blachly and Giordano~\cite{d1}, the impurity
concentration was such that, $\hbar/\tau_s^0\sim 0.1$~K, which is much
less than the Kondo temperature for Cu(Fe), $T_K \sim 3$~K. Therefore,
the second of Eqs.~(\ref{final}) should apply.
Figure~\ref{fig:comp} shows a comparison between the experimental data
of Blachly and Giordano\cite{d1}
and the theoretical predictions.  Each black square corresponds to one
sample.  The best fit to the data 
was obtained with $\tau_s^0= 1.3\times 10^{-10}$~s, whereas experimentally
the spin scattering time is on the order of $10\times 10^{-11}$.  This 
discrepancy also persists for the Cu(Mn) alloys for thicknesses
 of order $750-400$\AA.
However, for the thinnest Cu(Mn) alloys\cite{d3}  
Jacobs and Giordano have shown that excellent agreement exists between
theory and experiment for $\tau_s^o=6\times 10^{-11}$, which is well
within the experimental uncertainty of the measured value,
$\tau_s^o=6\times 10^{-11}$.   

While theory and experiment are in good agreement for thin samples,
there is a key experimental ambiguity that
surrounds these results, namely is there a well-defined Kondo temperature
for the thinnest samples and for those with mean free paths of order
$250$\AA.  This question is most relevant in light of the experiments
of Yanson and colleagues\cite{yanson} who have shown that in point contacts,
huge fluctuations in the Kondo temperature occur for contact diameters
of order $100$\AA.  For an inhomogeneous system, the
density of states becomes position dependent:  $\ro(x) = \ro + \delta \ro(x)$.
Consequently, the position-dependent Kondo temperature will be
\beq
T_K(x) = \eps_F \exp[{1/2\ro(x) J}]\approx \eps_F \exp[{1/2\ro(x) J}] = T_K
 \exp[-\delta \ro(x)/2\ro^2 J].
\eeq
Typically $\ro J \sim 0.1$, which means that even a 10\% change in
the density of states can produce 100\% change in the Kondo
temperature.  The effect is even stronger for alloys with lower
Kondo temperatures.  From elementary scattering theory,
$\delta\rho_0(x)=\rho_01/\sqrt{k_F\lambda}$, where $\lambda$ is 
the smaller of the mean-free path and the sample thickness.  For
$\ell=100$\AA, $\delta\rho_0(x)/\rho_0=.1$.  Hence, we expect a
100$\%$ change in the Kondo temperature for such samples.  Those impurities
having high Kondo temperatures will not contribute to the Kondo resistivity.
Consequently, fluctuations in the density of states can effectively decrease
the concentration of active spin-flip scattering centers that could contribute
to the Kondo logarithm. This will lead to
an enhancement in the spin-scattering rate over the bulk value.
Hence, the question as to how well-defined the Kondo temperature is in
the thinnest and most disordered samples should be resolved before a complete
experimental understanding of the Kondo effect in dirty alloys can be reached.

\begin{acknowledgments}
This work was funded by the DMR of the NSF.
\end{acknowledgments}


\begin{thebibliography}{99}
\bibitem{s1} G. Chen and N. Giordano, Physica B {\bf 165 \& 166}, 455 (1990).
\bibitem{s2} G. Chen and N. Giordano, Phys. Rev. Lett. {\bf 66}, 209 (1991).
\bibitem{s3} J. F. DiTusa, K. Lin, M. Park, M. S. Isaccson, and J. M. Parpia,
Phys. Rev. Lett. {\bf 68}, 1156 (1992).
\bibitem{s4}V. Chandrasekhar, P. Santhanam, N. A. Penebre, R. A. Webb,
H. Vloeberghs, C. Van Haesendonck, and Y. Bruynseraede, Phys. Rev.
Lett. {\bf 72}, 2053 (1994).
\bibitem{s5}O. Ujsaghy, A. Zawadowski, and B. L. Gyorffy, Phys. Rev.
Lett. {\bf 76}, 2378 (1996); O. Ujsaghy and A. Zawadowski,
 Phys. Rev. B {\bf 57}, 11598 (1998).
\bibitem{d1}M. A. Blachly 
and N. Giordano, Phys. Rev B {\bf 51},
12537 (1995); ibid, Europhys. Lett. {\bf 27}, 687 (1995).
\bibitem{d2} T. M. Jacobs and N. Giordano, Europhys. Lett. {\bf 44},
74 (1998).
\bibitem{d3}T. M. Jacobs and N. Giordano, this issue.
\bibitem{yanson}I. K. Yanson, V. V. Fisun, R. Hesper, A. V. Khotkevich,
J. M. Krans, J. A. Mydosh, and J. M van Ruitenbeek, Phys. Rev. Lett. 
{\bf 74}, 302 (1995).
\bibitem{ivar}I. Martin, Y. Wan, and P. Phillips, Phys. Rev. Lett. {\bf 78},
114 (1997).
\bibitem{everts} H. U. Everts and J. Keller, Z. Phys. {\bf 240}, 281 (1970).
\bibitem{bf}K.-P. Bohnen and K. H. Fischer, J. Low Temp. Phys. {\bf 12}, 559 (1970).
\bibitem{ohk}F. J. Ohkawa and H. Fukuyama,
 J. Phys. Soc. Jpn. {\bf 55},
2640 (1984); For a review see F. J. Ohkawa, Prog. Theor. Phys. Suppl.
{\bf 84}, 166 (1985).
\bibitem{vladar}K. Vladar and G. T. Zimanyi, J. Phys. C {\bf 18},
3739 (1985).
\bibitem{alt}B. L. Altshuler, et. al. {\bf Quantum Theory of Solids}, Ed. I. M. Lifshits,
(MIR, Moscow), 1982.
\bibitem{abr1965}
A. A. Abrikosov, Physics {\bf 2}, 5 (1965). 
\bibitem{chak}S. Chakravarty and C. Nayak, cond-mat/9911441.
\bibitem{volkov}
V. A. Volkov, JETP Lett. {\bf 36}, 475 (1982).
\end{thebibliography}
\end{document}